\documentclass[amssymb,aps,prl,twocolumn,groupedaddress,showpacs]{revtex4}

\usepackage{graphicx}

\newcommand{\C}{{\cal C}}

\newcommand{\K}{{\cal K}}
\renewcommand{\L}{{\cal L}}
\newcommand{\M}{{\cal M}}
\renewcommand{\P}{{\cal P}}
\renewcommand{\S}{{\cal S}}
\newcommand{\p}{\varphi}

\begin{document}

\title{Complete Determination of the Spectrum of a Transfer Operator associated with Intermittency}

\author{Thomas Prellberg}
\email{thomas.prellberg@tu-clausthal.de}
\affiliation{Institut f\"ur Theoretische Physik, Technische Universit\"at Clausthal, Arnold-Sommerfeld Stra\ss e 6, D-38678 Clausthal-Zellerfeld, Germany}

\date{\today}

\begin{abstract}
It is well established that the physical phenomenon of intermittency can be investigated via the spectral analysis of a 
transfer operator associated with the dynamics of an interval map with indifferent fixed point. We present here for the 
first time a complete spectral analysis for an example of such an intermittent map, the Farey map. We give a simple proof 
that the transfer operator is self-adjoint on a suitably defined Hilbert space and show that its spectrum decomposes into
a continuous part (the interval $[0,1]$) and isolated eigenvalues of finite multiplicity. Using a suitable first-return map, 
we present a highly efficient numerical method for the determination of all the eigenvalues, including the ones embedded in 
the continuous spectrum.
\end{abstract}

\pacs{02.30.Sa, 02.30.Tb, 02.70.Hm, 05.10.-a, 05.45.Ac}

\maketitle

\section{\label{intro}Introduction}

Intermittency, one of the main routes from order to chaos \cite{schuster1995a-a}, is characterized by the loss of stability of a fixed point 
of the dynamics. The dynamics directly at the transition is determined by a marginally unstable fixed point near which 
trajectories are slowed down severely, leading to the for intermittency characteristic interplay of chaotic and regular dynamics.
Such a behavior can easily be modeled by a map $f$ of the unit interval $[0,1]$ which is uniformly expanding everywhere except 
near an indifferent fixed point at zero, where $f(x)\sim x+cx^{r+1}$ as $x\rightarrow 0$ with exponent $1+r>1$.
A typical example is the Manneville map $f(x)=x+x^{1+r}\mbox{mod} 1$ \cite{manneville1980a-a}.

There have been many theoretical approaches to the description of intermittency such as renormalization group analysis \cite{hu1982a-a}. 
An approach suited to rigorous treatment is given by the thermodynamic formalism \cite{ruelle1978a-a} and leads to the spectral analysis 
of transfer operators. In contrast with uniformly expanding maps, however, the indifferent fixed point induces
non-Gibbsian equilibrium states or, more precisely, weakly Gibbsian states \cite{maes2000a-a}. 
Therefore, deeper understanding based on rigorous analysis has been hard 
to come by.
Some progress was made by studying a suitably defined 
piecewise linear interval map \cite{gaspard1988a-a}, albeit at the cost of simplifying the dynamics by severely reducing correlations. 
In \cite{prellberg1991a-a,prellberg1992a-a} we argued the disappearance of a spectral gap for the Perron-Frobenius operator, implying 
loss of an exponential decay of correlations for the dynamics. The decay of correlations has later been shown to follow a power law; 
in the analytic case ($r=1$) a numerical estimate of $t^{-2}$ \cite{lambert1993a-a} has been complemented by a rigorous upper bound 
of $t^{-2}\log t$, obtained by random perturbation techniques \cite{liverani1999a-a}. Only very recently has the Perron-Frobenius 
operator for a particular intermittent map, the Farey map, been shown to be self-adjoint on an appropriate Hilbert 
space \cite{isola2001a-a} with continuous spectrum on the interval $[0,1]$. There is also numerical work available describing 
just how the continuous spectrum emerges when approaching the intermittency transition \cite{kaufmann1996a-a}. 

Much less is known 
about the spectrum of the Ruelle-Perron-Frobenius (RPF) operator, which generalizes the Perron-Frobenius operator within the framework 
of the thermodynamic formalism. In \cite{prellberg1991a-a,prellberg1992a-a} we showed that for a general class of intermittent maps
this operator is quasi-compact with essential spectral radius equal to one, and that the leading eigenvalue undergoes a phase transition
characteristic for intermittent dynamics. Only recently was it shown that for a class of piecewise analytic maps the continuous spectrum 
is restricted to the interval $[0,1]$ on an appropriate function space \cite{rugh1999a-a}. 
The present work gives for the first time a complete spectral analysis of the RPF operator for 
an intermittent map, the Farey map.

A promising conceptual approach to the study of intermittency is the study of a first-return map (or induced map) with respect to a domain 
of phase space away from the intermittent region. In a pioneering study \cite{prellberg1991a-a,prellberg1992a-a} we introduced a suitably 
modified transfer operator for the 
induced map and related its spectral properties to the ones of the transfer operator of the intermittent system. This approach
has since been extended to the study of other quantities such as regularized Fredholm determinants and dynamical zeta functions
\cite{dodds1993a-a,dodds1993a-b,isola1995a-a,isola2001a-a} and multi-dimensional systems \cite{pollicott2001a-a}.
One essential idea employed is that one can understand a general transfer operator $\P$ of an intermittent map by first 
considering its ``intermittent part'' $\P_0$ and then viewing the ``chaotic remainder'' $\P_1=1-\P$ as a perturbation. 
In this way, many of the results presented here can in principle be extended to more general intermittent maps, although we 
shall focus our attention on the {\it Farey map} of the interval $[0,1]$ onto itself, which is defined as
\begin{equation}
f(x)=\left\{\begin{array}{c}
f_0(x)=x/(1-x)\;,\quad\mbox{if $0\leq x\leq1/2$,}\\
f_1(x)=(1-x)/x\;,\quad\mbox{if $1/2<x\leq1$.}
\end{array}\right.
\end{equation}
We denote the inverses by $F_0(x)={f_0}^{-1}(x)=x/(1+x)$ and $F_1(x)={f_1}^{-1}(x)=1/(1+x)$.

The advantage of this map is that each branch can be analytically extended to the whole Riemann sphere and that higher iterates of the 
left branch can be given exactly, ${f_0}^n(x)=x/(1-nx)$, a fact that is intimately connected to the global conjugacy of $f_0(x)$
to the shift $x\rightarrow x-1$ \footnote{For general parabolic fixed points such a conjugacy can always be found 
locally, a fact that is used in the analysis of \cite{rugh1999a-a}.}. The thermodynamic formalism suggests the study of the 
RPF operator $\P$ associated with a map $f$, which is given by 
$\P\p(x)=\sum_{f(y)=x}|f'(y)|^{-\beta}\p(f(x))$ with $\beta\in\mathbb{R}$. For the Farey map the operator consists of two terms which 
can be readily identified with the ``intermittent'' and ``chaotic'' parts. We thus write $\P=\P_0+\P_1$ where 
$\P_i\p(x)=|{F_i}'(x)|^\beta\p(F_i(x))$ and $\beta\in\mathbb{R}$, i.e.
\begin{equation}
\P\p(x)=\frac1{(1+x)^{2\beta}}\left[\p\left(\frac x{1+x}\right)+\p\left(\frac1{1+x}\right)\right]\;.
\end{equation}

Despite the apparent simplicity of this operator, its properties have not been well understood until recently. In this paper we first present 
an abstract analysis describing its spectrum and then proceed to determine all of its eigenvalues with a highly efficient numerical method
based on the first-return map.

\section{\label{spectrum}Spectral Properties of the Transfer Operator}

As indicated above, one can understand the spectral properties of $\P$ by first studying the operator $\P_0$ and then viewing $\P_1$ as
a perturbation. 

Under conjugacy with $\C\p(x)=\p(f_1(x))$, the operator $\P_0$ transforms to the shift operator $S\p(x)=\p(1+x)$.
This operator is now independent of $\beta$, and on a suitably defined function space has continuous spectrum $\sigma(S)=\sigma_c(S)=[0,1]$. 
This can be understood by considering functions $\p$ which are obtained by a generalized Laplace transform 
$\p(x)=\L\psi(x)=\int_0^\infty\exp(-sx)\psi(s)d\mu(s)$ of square integrable functions $\psi\in L^2(\mathbb{R}_+,\mu)$.
(This automatically ensures analyticity of $\p(x)$ in a suitable domain.) The action of the operator $S$ is then conjugate to multiplication
by $\exp(-s)$ on $L^2(\mathbb{R}_+,\mu)$. The spectrum of this multiplication operator is continuous and given by the closure of the range of 
the multiplying function. This proves that $\P_0$ is a bounded self-adjoint operator on the Hilbert space $\C\L L^2(\mathbb{R}_+,\mu)$ with 
spectrum $\sigma(\P_0)=\sigma_c(\P_0)=[0,1]$.

To study the spectrum of $\P$, we consider the identity 
\begin{equation}
\label{identity}
1-z\P=(1-z\P_0)(1-\M_z)
\end{equation}
with 
$\M_z=(1-z\P_0)^{-1}z\P_1$ being an operator-valued analytic function for $z\in\mathbb{C}-[1,\infty)$ \footnote{In \cite{isola2001a-a} a 
related identity has been used, leading to the study of $\M_z'=\P_1(1-z\P_0)^{-1}$. Obviously, $\M_z$ and $\M_z'$ are related by conjugacy.}
We point out that (\ref{identity}) remains valid when $\M_z$ is analytically continued across the cut $[1,\infty)$.
Expanding formally in powers of $z$, we find that $\M_z=\sum_{n=1}^\infty z^n{\P_0}^{n-1}\P_1$ is a transfer operator associated with the 
induced map $g$ on $[1/2,1]$ with the branches $g_n={f_0}^{n-1}f_1$ for $n\in\mathbb{N}$, where the return time $n$ has been encoded via
a multiplicative weight factor $z^n$.

To determine the spectral properties of the operator $\M_z$ we note that the induced map $g$ on $[1/2,1]$ is expanding. It follows that
for $|z|<1$ the operator $\M_z$ acting on a Frechet space of functions analytic in a domain $\Omega$ containing the interval $[1/2,1]$ 
is nuclear of order zero. Therefore, the analytic continuation of $\M_z$ is also nuclear of order zero 
(which can be seen via the analytic continuation of the associated Fredholm determinants, see \cite{rugh1999a-a}).

The identity (\ref{identity}) immediately implies that $\lambda=z^{-1}$ is an eigenvalue of $\P$ if and only if $1$ is an eigenvalue of
$\M_z$. The respective eigenspaces are identical, so that the geometric multiplicities of the respective eigenvalues $z^{-1}$ and $1$ are 
the same. Using the analytic continuation of $\M_z$ across the cut $[1,\infty)$, general analyticity arguments and the nuclearity of $\M_z$ 
imply that the non-zero point spectrum of $\P$ consists of isolated eigenvalues with finite multiplicity, with $0$ and $1$ as the only possible
accumulation points. 
Moreover, bounds on $\M_z$ for $|z|<1$ imply that there are only finitely many eigenvalues with modulus greater than one. 
Along the cut $z\in(1,\infty)$ we see that 
if the analytic continuation of $\M_z$ does not have an eigenvalue $1$ then $1-\M_z$ is invertible, implying 
that $\sigma_c(\P)=\sigma_c(\P_0)=[0,1]$, albeit with the possibility of embedded eigenvalues. Similar arguments can be found 
in a rather general setting in \cite{rugh1999a-a}. We further know that zero is an eigenvalue of infinite multiplicity, as 
$\P\p=0$ for any $\p$ satisfying $\p(x)=-\p(1-x)$.

We now show that $\P$ is in fact self-adjoint on a suitably defined Hilbert space for which we can give a representation of
$\P$ which is explicitly symmetric. For this, we consider the transfer operator $Q$ for a general M\"obius transformation 
$h(x)=(ax+b)/(cx+d)$ with $a,b,c,d\in\mathbb{R}$ and $\sigma=ad-bc=\pm1$. We have $Q\p(x)=(a-cx)^{-2\beta}\p(\frac{dx-b}{a-cx})$ 
and under a generalized Laplace transform $Q$ is conjugated to an integral operator $\K\psi(s)=\int_0^\infty K(s,t)\psi(t)d\mu(t)$ on
$L^2(\mathbb{R}_+,\mu)$. Using the Schl\"afli integral representation for Bessel functions \cite{arfken2001a-a}, one can calculate for
$c\neq0$ directly
\begin{equation}
\mu(s)K(s,t)=
\frac1c\exp\frac{as+td}c\left(\frac st\right)^{\beta-\frac12}
Z_{2\beta-1}(\frac2c\sqrt{st})
\end{equation}
with $Z_\nu(u)=I_\nu(u)$ for $\sigma=1$ and $Z_\nu(u)=J_\nu(u)$ for $\sigma=-1$.
Demanding that the kernel be symmetric determines the appropriate measure. In this way, we obtain for $\P_1$ the measure
$\mu(s)=e^{-s}s^{2\beta-1}$ and the kernel $K(s,t)=(st)^{\frac12-\beta}J_{2\beta-1}(2\sqrt{st})$. Combining this with our discussion
of the operator $\P_0$, we get a conjugacy of the transfer operator $\P$ to the self-adjoint operator 
\begin{equation}
\label{selfadj}
\S\psi(s)=e^{-s}\psi(s)+\int_0^\infty K(s,t)\psi(t)d\mu(t)
\end{equation}
Expressed differently, $\P$ is self-adjoint on the Hilbert space
\begin{equation}
\left\{\p(x)=x^{-2\beta}\int_0^\infty e^{-s\frac{1-x}x}\psi(s)d\mu(s):\psi\in L^2(\mathbb{R}_+,\mu)\right\}\nonumber
\end{equation}
with $\mu(s)=e^{-s}s^{2\beta-1}$ and appropriate induced inner product. $\M_z$ also leaves this Hilbert space invariant, and via a
similar argument one can show self-adjointness for real $z<1$.

The case $\beta=1$ has already been treated in \cite{isola2001a-a}, although with a slightly different choice of measure, motivated by 
the Hilbert space approach for the continued fraction transform \cite{mayer1987a-a,mayer1990a-a}. There, it was shown that for 
$\tilde\mu(s)=s/(\exp s-1)$ the only non-zero eigenvalue of $\P$ is $1$ with eigenfunction $\p(x)=1/x$. The measure 
$\mu(s)=s e^{-s}$ considered here has the advantage that the self-adjointness is explicitly evident from (\ref{selfadj}). 
However, $\p(x)=1/x$ is not an element of
the Hilbert space considered here, as it corresponds to $\psi(s)=1/s$ which has infinite norm in $L^2(\mathbb{R}_+,\mu)$.
We find this quite natural,
as the usual interpretation of this eigenfunction is as an absolutely continuous invariant density with respect to the Lebesgue measure on the
interval $[0,1]$, which in this case is non-normalizable.

\section{\label{numerics}Numerical Analysis of the Spectrum}

As shown above, $\M_z$ is an operator-valued analytic function with nuclear spectrum. By standard analytic perturbation 
theory \cite{kato1980a-a}, the eigenvalues of $\M_z$ are (branches of) analytic functions in $z$ with only algebraic singularities. 
This therefore provides a possibility to compute the eigenvalues of $\P$ numerically; choosing an 
eigenvalue branch $\lambda_n(z)$, one simply needs to solve $\lambda_n(z)=1$ to get an eigenvalue $\lambda=1/z$ of $\P$.
The essential advantage over working
directly with $\P$ is that we have removed the continuous spectrum which presents enormous difficulties for a direct numerical analysis.

As the operator $\M_z$ acts on a Hilbert space of analytic functions, it is reasonable to consider the action of $\M_z$ on 
coefficients of power series. From the explicit expansion 
$
\M_z\p(x)=\sum_{n=1}^\infty z^n(1+nx)^{-2\beta}\p(1-x/(1+nx))
$
we obtain by expanding $\p(x)$ in a power series around $x=1$ matrix elements $\M_z^{n,m}$ in terms of the polylogarithm 
$\mbox{Li}_s(z)=\sum_{n=1}^\infty\frac{z^n}{n^s}$ as
\begin{eqnarray}
M_z^{n,m}&=&\sum_{k=0}^n(-1)^{n-k}{-2\beta-m\choose k}{-2\beta-k\choose n-k}\nonumber\\
&&\times\left(\frac1z\mbox{Li}_{2\beta+m+k}(z)-1\right)\;.
\end{eqnarray}
We then approximate $\M_z$ with truncated operators $\M_z^{(N)}$ acting on a subspace of polynomials of at most degree $N$. 
It turns that this approximation works very well and that one obtains the values of the leading eigenvalues to a high accuracy.  In this
setting, this idea goes back to \cite{dodds1993a-a}.  In \cite{daude1997a-a}, where a related operator was studied, 
the leading eigenvalues have been obtained in this way with accuracy of $10^{-25}$ \footnote{As in in \cite{daude1997a-a}, 
we find that we can improve convergence by expanding around a different point. However, this leads to the introduction of spurious 
eigenvalues which make the analysis more difficult. By choosing an expansion around $x=1$ we avoid spurious eigenvalues and find 
that, overall, the spectrum is better approximated.}.

\begin{figure}
\includegraphics[width=8cm]{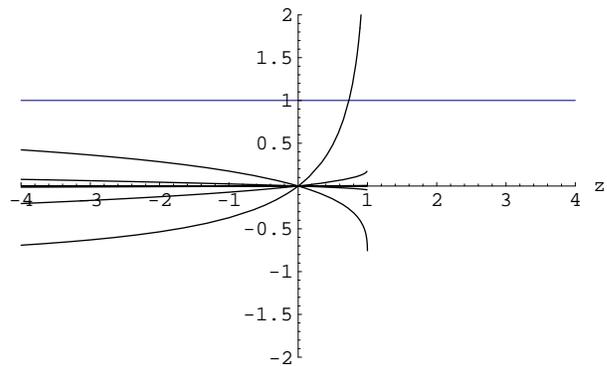}
\caption{\label{fig1} The seven leading eigenvalues of $\M_z$ for $\beta=0.5$. The analytic continuation of the eigenvalue branches beyond
$z=1$ is complex valued and has a jump across the cut $[1,\infty)$. Only the largest eigenvalue branch intersects $\lambda=1$.}
\end{figure}
\begin{figure}
\includegraphics[width=8cm]{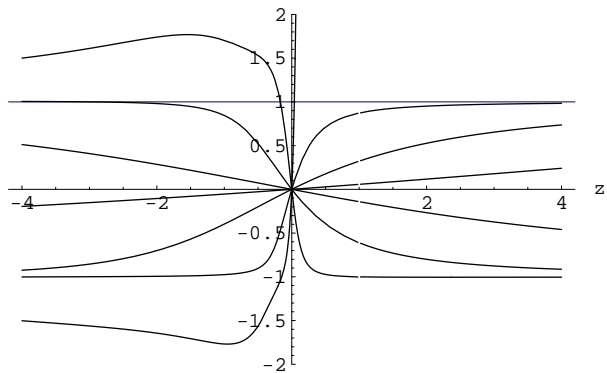}
\caption{\label{fig2} The seven leading eigenvalues of $\M_z$ for $\beta=-3$. Here, the eigenvalue branches can be continued beyond $z=1$
and intersect $\lambda=1$ at $z=-2.971$, $z=-0.168$, $z=0.038$, and $z=13.101$, the last value being outside the range of this plot.}
\end{figure}
\begin{figure}
\includegraphics[width=8cm]{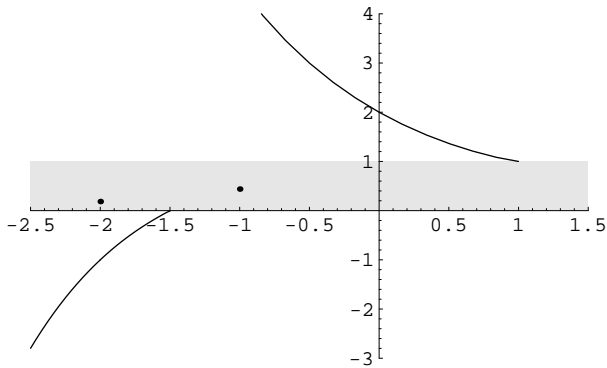}
\caption{\label{fig3} The spectrum of $\P$ as a function of $\beta$, showing two non-zero eigenvalue branches and
isolated eigenvalues at $\beta=-1$ and $\beta=-2$ which are embedded in the continuous spectrum $[0,1]$.}
\end{figure}

We begin the description of the results of our analysis by briefly considering the special cases $\beta=-N/2$ with $N\in\mathbb{N}_0$. 
For these values of $\beta$ we find that polynomials of at most degree $N$ give an $(N+1)$-dimensional invariant subspace for 
$\M_z$ and $\P$. The corresponding truncated matrix $\M_z^{(N)}$ has entries which are rational functions in $z$ from which one can easily
calculate $N+1$ eigenvalues exactly.  
Numerically we observe that eigenfunctions of $\M_z$ which are not in this invariant subspace have eigenvalues which are strictly 
smaller in modulus.
It is especially noteworthy that these leading eigenvalues can be 
analytically continued across $z=1$, which allows for eigenvalues of the Farey operator $\P$ embedded in the 
continuous spectrum. However, a more detailed numerical analysis shows that this analytic continuation is possible {\it only} for 
$\beta=-N/2$, and a small deviation from these values leads to eigenvalue branches $\lambda_n(z)$ whose analytic 
extension generically shows a non-vanishing jump in the imaginary part along the cut $[1,\infty)$ \footnote{One can understand this 
by considering the analytic extension of $\mbox{Li}_s(z)$, which across the cut $[1,\infty)$ jumps at $x>1$ by an amount
$2\pi i\log^{s-1}(x)/\Gamma(s)$.}.
Therefore $\lambda_n(z)=1$ for real $z>1$ can only be satisfied when $\beta=-N/2$, and no eigenvalues of $\P$ embedded in the continuous
spectrum exist for other values of $\beta$. Thus, in what follows we shall mainly be concerned with real $z\leq1$.

For $\beta>1$ we find that all the eigenvalues of $\M_z$ are strictly less than one in modulus, so that $\P$ has continuous spectrum 
$[0,1]$ and no non-zero eigenvalues at all. As $\beta$ decreases the eigenvalues of $\M_z$ increase in value. For $\beta>1$, we find
that the leading eigenvalue branch of $\M_z$ intersects $\lambda=1$, leading to the emergence of a simple leading eigenvalue of $\P$ 
for $\beta<1$ \footnote{In \cite{prellberg1991a-a,prellberg1992a-a} it was 
proved that this leading eigenvalue decreases to $1$ like $-(1-\beta)/\log(1-\beta)$ as $\beta$ approaches $1$ from below.}.
The $z$-dependence of the spectrum of $\M_z$ is shown in Figure \ref{fig1} for $\beta=1/2$.  Only the largest eigenvalue intersects
$\lambda=1$, implying that $\P$ has only one non-zero eigenvalue.
At $\beta=-3/2$ a second eigenvalue branch begins to intersect 
$\lambda=1$ at large negative $z$, implying that $\P$ has two non-zero eigenvalues. The second eigenvalue of $\P$ is negative and 
becomes in modulus equal to the essential spectral radius at $\beta=-2$, and therefore determines for $\beta<-2$ the spectral
gap which controls the decay of correlations. 
Figure \ref{fig2} illustrates the leading spectrum of $\M_z$ for $\beta=-3$. At this special value of $\beta$
we can extend the eigenvalue branches beyond $z=1$ and find a total of four eigenvalue branches which cross $\lambda=1$. One of these 
crossings is at $z=13.101$, corresponding to an eigenvalue of $\P$ embedded in the continuous spectrum.
Proceeding in this fashion, we can numerically determine the complete spectrum of $\P$ for arbitrary real $\beta$. Figure \ref{fig3}
shows the spectrum obtained in this way for $-2.5\leq\beta\leq1.5$. 

In summary, we have abstractly characterized the spectrum of the transfer operator for the Farey map and presented a highly 
efficient method for the explicit computation of its eigenvalues. Of special interest is the interplay of 
sub-dominant eigenvalues and the continuous spectrum. Both the rigorous arguments and numerical methods are generalizable to more 
complicated systems with intermittency. 


\begin{thebibliography}{22}
\expandafter\ifx\csname natexlab\endcsname\relax\def\natexlab#1{#1}\fi
\expandafter\ifx\csname bibnamefont\endcsname\relax
  \def\bibnamefont#1{#1}\fi
\expandafter\ifx\csname bibfnamefont\endcsname\relax
  \def\bibfnamefont#1{#1}\fi
\expandafter\ifx\csname citenamefont\endcsname\relax
  \def\citenamefont#1{#1}\fi
\expandafter\ifx\csname url\endcsname\relax
  \def\url#1{\texttt{#1}}\fi
\expandafter\ifx\csname urlprefix\endcsname\relax\def\urlprefix{URL }\fi
\providecommand{\bibinfo}[2]{#2}
\providecommand{\eprint}[2][]{\url{#2}}

\bibitem[{\citenamefont{Schuster}(1995)}]{schuster1995a-a}
\bibinfo{author}{\bibfnamefont{H.~G.} \bibnamefont{Schuster}},
  \emph{\bibinfo{title}{Deterministic Chaos, An Introduction}}
  (\bibinfo{publisher}{Physik-Verlag}, \bibinfo{address}{Weinheim},
  \bibinfo{year}{1995}).

\bibitem[{\citenamefont{Manneville}(1980)}]{manneville1980a-a}
\bibinfo{author}{\bibfnamefont{P.}~\bibnamefont{Manneville}},
  \bibinfo{journal}{J. Physique} \textbf{\bibinfo{volume}{41}},
  \bibinfo{pages}{1235} (\bibinfo{year}{1980}).

\bibitem[{\citenamefont{Hu and Rudnick}(1982)}]{hu1982a-a}
\bibinfo{author}{\bibfnamefont{B.}~\bibnamefont{Hu}} \bibnamefont{and}
  \bibinfo{author}{\bibfnamefont{J.}~\bibnamefont{Rudnick}},
  \bibinfo{journal}{Phys. Rev. Lett.} \textbf{\bibinfo{volume}{48}},
  \bibinfo{pages}{1645} (\bibinfo{year}{1982}).

\bibitem[{\citenamefont{Ruelle}(1978)}]{ruelle1978a-a}
\bibinfo{author}{\bibfnamefont{D.}~\bibnamefont{Ruelle}},
  \emph{\bibinfo{title}{Thermodynamic Formalism}}
  (\bibinfo{publisher}{Addison-Wesley}, \bibinfo{year}{1978}).

\bibitem[{\citenamefont{Maes et~al.}(2000)\citenamefont{Maes, Redig, Takens,
  van Moffaert, and Verbitsky}}]{maes2000a-a}
\bibinfo{author}{\bibfnamefont{C.}~\bibnamefont{Maes}},
  \bibinfo{author}{\bibfnamefont{F.}~\bibnamefont{Redig}},
  \bibinfo{author}{\bibfnamefont{F.}~\bibnamefont{Takens}},
  \bibinfo{author}{\bibfnamefont{A.}~\bibnamefont{van Moffaert}},
  \bibnamefont{and}
  \bibinfo{author}{\bibfnamefont{E.}~\bibnamefont{Verbitsky}},
  \bibinfo{journal}{Nonlinearity} \textbf{\bibinfo{volume}{13}},
  \bibinfo{pages}{1681} (\bibinfo{year}{2000}).

\bibitem[{\citenamefont{Gaspard and Wang}(1988)}]{gaspard1988a-a}
\bibinfo{author}{\bibfnamefont{P.}~\bibnamefont{Gaspard}} \bibnamefont{and}
  \bibinfo{author}{\bibfnamefont{X.~J.} \bibnamefont{Wang}},
  \bibinfo{journal}{Proc. Natl. Acad. Sci. USA} \textbf{\bibinfo{volume}{85}},
  \bibinfo{pages}{4591} (\bibinfo{year}{1988}).

\bibitem[{\citenamefont{Prellberg}(1991)}]{prellberg1991a-a}
\bibinfo{author}{\bibfnamefont{T.}~\bibnamefont{Prellberg}}, Ph.D. thesis,
  \bibinfo{school}{Virginia Tech} (\bibinfo{year}{1991}).

\bibitem[{\citenamefont{Prellberg and Slawny}(1992)}]{prellberg1992a-a}
\bibinfo{author}{\bibfnamefont{T.}~\bibnamefont{Prellberg}} \bibnamefont{and}
  \bibinfo{author}{\bibfnamefont{J.}~\bibnamefont{Slawny}},
  \bibinfo{journal}{J. Stat. Phys.} \textbf{\bibinfo{volume}{66}},
  \bibinfo{pages}{503} (\bibinfo{year}{1992}).

\bibitem[{\citenamefont{Lambert et~al.}(1993)\citenamefont{Lambert, Siboni, and
  Vaienti}}]{lambert1993a-a}
\bibinfo{author}{\bibfnamefont{A.}~\bibnamefont{Lambert}},
  \bibinfo{author}{\bibfnamefont{S.}~\bibnamefont{Siboni}}, \bibnamefont{and}
  \bibinfo{author}{\bibfnamefont{S.}~\bibnamefont{Vaienti}},
  \bibinfo{journal}{J. Stat. Phys.} \textbf{\bibinfo{volume}{72}},
  \bibinfo{pages}{1305} (\bibinfo{year}{1993}).

\bibitem[{\citenamefont{Liverani et~al.}(1999)\citenamefont{Liverani, Saussol,
  and Vaienti}}]{liverani1999a-a}
\bibinfo{author}{\bibfnamefont{C.}~\bibnamefont{Liverani}},
  \bibinfo{author}{\bibfnamefont{B.}~\bibnamefont{Saussol}}, \bibnamefont{and}
  \bibinfo{author}{\bibfnamefont{S.}~\bibnamefont{Vaienti}},
  \bibinfo{journal}{Ergodic Theory Dynam. Systems}
  \textbf{\bibinfo{volume}{19}}, \bibinfo{pages}{671} (\bibinfo{year}{1999}).

\bibitem[{\citenamefont{Isola}({\natexlab{a}})}]{isola2001a-a}
\bibinfo{author}{\bibfnamefont{S.}~\bibnamefont{Isola}},
  \bibinfo{note}{preprint, 2001}.

\bibitem[{\citenamefont{Kaufmann et~al.}(1996)\citenamefont{Kaufmann, Lustfeld,
  and Bene}}]{kaufmann1996a-a}
\bibinfo{author}{\bibfnamefont{Z.}~\bibnamefont{Kaufmann}},
  \bibinfo{author}{\bibfnamefont{H.}~\bibnamefont{Lustfeld}}, \bibnamefont{and}
  \bibinfo{author}{\bibfnamefont{J.}~\bibnamefont{Bene}},
  \bibinfo{journal}{Phys. Rev. E} \textbf{\bibinfo{volume}{53}},
  \bibinfo{pages}{1416} (\bibinfo{year}{1996}).

\bibitem[{\citenamefont{Rugh}(1999)}]{rugh1999a-a}
\bibinfo{author}{\bibfnamefont{H.~H.} \bibnamefont{Rugh}},
  \bibinfo{journal}{Invent. Math.} \textbf{\bibinfo{volume}{135}},
  \bibinfo{pages}{1} (\bibinfo{year}{1999}).

\bibitem[{\citenamefont{Dodds}()}]{dodds1993a-a}
\bibinfo{author}{\bibfnamefont{P.}~\bibnamefont{Dodds}},
  \bibinfo{note}{{H}onour's thesis, The University of Melbourne (1993)}.

\bibitem[{\citenamefont{Dodds}(1993)}]{dodds1993a-b}
\bibinfo{author}{\bibfnamefont{P.}~\bibnamefont{Dodds}}, Master's thesis,
  \bibinfo{school}{The University of Melbourne} (\bibinfo{year}{1993}).

\bibitem[{\citenamefont{Isola}({\natexlab{b}})}]{isola1995a-a}
\bibinfo{author}{\bibfnamefont{S.}~\bibnamefont{Isola}},
  \bibinfo{note}{preprint, 1995}.

\bibitem[{\citenamefont{Pollicott and Yuri}(2001)}]{pollicott2001a-a}
\bibinfo{author}{\bibfnamefont{M.}~\bibnamefont{Pollicott}} \bibnamefont{and}
  \bibinfo{author}{\bibfnamefont{M.}~\bibnamefont{Yuri}},
  \bibinfo{journal}{Nonlinearity} \textbf{\bibinfo{volume}{14}},
  \bibinfo{pages}{1265} (\bibinfo{year}{2001}).

\bibitem[{\citenamefont{Arfken and Weber}(2001)}]{arfken2001a-a}
\bibinfo{author}{\bibfnamefont{G.~B.} \bibnamefont{Arfken}} \bibnamefont{and}
  \bibinfo{author}{\bibfnamefont{H.~J.} \bibnamefont{Weber}},
  \emph{\bibinfo{title}{Mathematical Methods for Physicists}}
  (\bibinfo{publisher}{Harcourt/Academic Press}, \bibinfo{year}{2001}).

\bibitem[{\citenamefont{Mayer and Roepstorff}(1987)}]{mayer1987a-a}
\bibinfo{author}{\bibfnamefont{D.~H.} \bibnamefont{Mayer}} \bibnamefont{and}
  \bibinfo{author}{\bibfnamefont{G.}~\bibnamefont{Roepstorff}},
  \bibinfo{journal}{J. Stat. Phys.} \textbf{\bibinfo{volume}{47}},
  \bibinfo{pages}{149} (\bibinfo{year}{1987}).

\bibitem[{\citenamefont{Mayer}(1990)}]{mayer1990a-a}
\bibinfo{author}{\bibfnamefont{D.~H.} \bibnamefont{Mayer}},
  \bibinfo{journal}{Commun. Math. Phys.} \textbf{\bibinfo{volume}{130}},
  \bibinfo{pages}{311} (\bibinfo{year}{1990}).

\bibitem[{\citenamefont{Kato}(1980)}]{kato1980a-a}
\bibinfo{author}{\bibfnamefont{T.}~\bibnamefont{Kato}},
  \emph{\bibinfo{title}{Perturbation Theory for Linear Operators}}
  (\bibinfo{publisher}{Springer}, \bibinfo{address}{Berlin},
  \bibinfo{year}{1980}).

\bibitem[{\citenamefont{Daud{\'e} et~al.}(1997)\citenamefont{Daud{\'e},
  Flajolet, and Vall{\'e}e}}]{daude1997a-a}
\bibinfo{author}{\bibfnamefont{H.}~\bibnamefont{Daud{\'e}}},
  \bibinfo{author}{\bibfnamefont{P.}~\bibnamefont{Flajolet}}, \bibnamefont{and}
  \bibinfo{author}{\bibfnamefont{B.}~\bibnamefont{Vall{\'e}e}},
  \bibinfo{journal}{Combin. Probab. Comput.} \textbf{\bibinfo{volume}{6}},
  \bibinfo{pages}{397} (\bibinfo{year}{1997}).

\end{thebibliography}

\end{document}